\documentclass[conference]{IEEEtran}
\usepackage[hyphens]{url}
\usepackage{hyperref}
\hypersetup{colorlinks,allcolors=blue}

\usepackage[backend=bibtex,style=ieee,citestyle=numeric-comp]{biblatex}
\usepackage{amsmath,amssymb,amsfonts}
\usepackage{algorithmic}
\usepackage{graphicx}
\usepackage{subcaption}
\usepackage{tabularx,booktabs}
\usepackage{textcomp}
\usepackage{xcolor}
\usepackage{orcidlink}
\usepackage{framed}
\usepackage{float}
\usepackage{enumitem} 
\def\BibTeX{{\rm B\kern-.05em{\sc i\kern-.025em b}\kern-.08em
    T\kern-.1667em\lower.7ex\hbox{E}\kern-.125emX}}
\graphicspath{{images/}}

\usepackage{caption}
\begin{document}

\title{Unsupervised Baseline Clustering and Incremental Adaptation for IoT Device Traffic Profiling}

\author{\IEEEauthorblockN{Sean M. Alderman\,\orcidlink{0009-0008-9913-1217}
}
\IEEEauthorblockA{\textit{The Beacom College of Computer \& Cyber Sciences} \\
\textit{Dakota State University}\\
Madison, SD, USA \\
sean.alderman@trojans.dsu.edu}
\and
\IEEEauthorblockN{John D. Hastings\,\orcidlink{0000-0003-0871-3622}
}
\IEEEauthorblockA{\textit{The Beacom College of Computer \& Cyber Sciences} \\
\textit{Dakota State University}\\
Madison, SD, USA \\
john.hastings@dsu.edu
}
}

\maketitle 

\begin{abstract}
The growth and heterogeneity of IoT devices create security challenges where static identification models can degrade as traffic evolves. This paper presents a two-stage, flow-feature-based pipeline for unsupervised IoT device traffic profiling and incremental model updating, evaluated on selected long-duration captures from the Deakin IoT dataset. For baseline profiling, density-based clustering (DBSCAN) isolates a substantial outlier portion of the data and produces the strongest alignment with ground-truth device labels among tested classical methods (NMI 0.78), outperforming centroid-based clustering on cluster purity. For incremental adaptation, we evaluate stream-oriented clustering approaches and find that BIRCH supports efficient updates (0.13 seconds per update) and forms comparatively coherent clusters for a held-out novel device (purity 0.87), but with limited capture of novel traffic (share 0.72) and a measurable trade-off in known-device accuracy after adaptation (0.71). Overall, the results highlight a practical trade-off between high-purity static profiling and the flexibility of incremental clustering for evolving IoT environments.
\end{abstract}


\section{Introduction}
The rapid proliferation and growing complexity of interconnected devices, especially in the Internet of Things (IoT) ecosystem, demand robust, adaptive systems for device identification and security. Modern wireless networks, marked by volatility and heterogeneity, challenge traditional algorithmic and heuristic methods, which often fail to deliver the timely, nuanced decisions essential for network management and threat mitigation. Consequently, recent research has increasingly adopted machine learning (ML) and deep learning (DL) techniques to analyze network traffic, typically captured as packet capture (PCAP) files, for accurate device classification and profiling.

Robust network device identification hinges on capturing both relatively stable device characteristics and changing behavioral patterns visible in traffic \cite{salman_machine_2022,ali_generic_2021}. Static fingerprints can be discriminative but may fail when devices update or exhibit context-dependent behavior, while purely behavioral approaches can be sensitive to noise and costly to maintain at scale \cite{salman_machine_2022,afifi_machine_2024,ali_generic_2021,wang_comprehensive_2024,hao_iottfid_2023,rabbani_lightweight_2024}. In dynamic IoT deployments, identification systems therefore benefit from packet-efficient features and incremental updating that can incorporate novel devices without retraining from scratch or degrading prior knowledge \cite{salman_machine_2022,afifi_machine_2024, hao_iottfid_2023,rabbani_lightweight_2024, kolcun_revisiting_2021, thom_smart_2022}. 

The dynamic nature of modern network environments renders static ML models inadequate, as their accuracy often degrades with new data \cite{afifi_machine_2024,hao_iottfid_2023,kolcun_revisiting_2021}. Online or incremental learning techniques, such as those used in IoTTFID \cite{hao_iottfid_2023}, enable efficient adaptation to new device types without resource-intensive retraining or catastrophic forgetting \cite{salman_machine_2022,rabbani_lightweight_2024,thom_smart_2022} supporting scalable profiling that balances computational and data requirements.

This work evaluates unsupervised clustering for baseline IoT device profiling and an incremental unsupervised approach for adaptive updates. The following research questions guide the study:

\begin{enumerate}[label={\textbf{RQ\arabic*:}},left=1.0em]
\item How well do packet-efficient flow features enable unsupervised IoT device profiling (cluster quality and noise rate)?

\item How effectively can an incremental unsupervised profiler adapt to new devices while retaining known-device performance and controlling update cost?
\end{enumerate}

\noindent In the process, this work contributes a packet-efficient profiling pipeline that benchmarks baseline clustering on long-duration real-world IoT traffic and evaluates incremental updates for novel-device adaptation with quality and cost metrics.

\section{Objectives and Contributions}
\label{objectives}

Guided by RQ1 and RQ2, this work tests the hypothesis that each device generates a distinctive, predictable fingerprint from its PCAP data, comprising two feature classes:
\begin{enumerate}
    \item Static and Unique Features: These include intrinsic identifiers such as MAC addresses, DHCP protocol data, and operating system indicators derived from packet headers, serving as essential but potentially spoofable reference points \cite{salman_machine_2022,hao_iottfid_2023,anaedevha_application_2024}.
    \item Dynamic and Behavioral Characteristics: These capture consistent network interactions, including protocol, port, and service usage, as well as patterns in activity cycles, packet sizes, and inter-arrival times. Such features are vital for distinguishing IoT devices with specialized functions \cite{ali_generic_2021,hu_network_2024,hao_iottfid_2023,kolcun_revisiting_2021,salman_machine_2022}.
\end{enumerate}

To address the limitations of static machine learning models, which lose accuracy with new device types or behavioral shifts, this study advocates online or incremental learning methods \cite{hao_iottfid_2023,kolcun_revisiting_2021}. These approaches enable continuous profile updates, efficiently incorporate new device classes without resource-intensive retraining, and mitigate catastrophic forgetting, i.e., degradation on previously learned device profiles as the model adapts \cite{salman_machine_2022,rabbani_lightweight_2024, thom_smart_2022}. By leveraging minimal packet data to classify devices based on their novelty, this methodology supports scalable profiling across diverse IoT ecosystems. By balancing generalization—clustering similar devices (e.g., same model or manufacturer) into logical classes—with recognition of each device's unique attributes, this methodology ensures a scalable, enduring system for device classification and network governance.

\section{Background, Motivation, \& Research Gaps}

This section establishes the background and motivation for our study and identifies the key research gaps that inform RQ1 and RQ2, focusing on unsupervised clustering for baseline profiling, incremental learning for adaptation under drift, and novelty conditions in long-duration IoT traffic.

The efficacy of unsupervised clustering models hinges on the quality of feature embeddings derived from raw PCAP data \cite{hamidouche_enhancing_2024,rabbani_lightweight_2024}. While older datasets like UNSW \cite{sivanathan_classifying_2019} have been the standard, the recently released Deakin IoT Traffic (D-IoT) dataset \cite{pasquini_descriptor_2025} offers a significant advancement for testing these models. Spanning 119 days with 110 million packets, D-IoT provides a high-fidelity environment for evaluating cluster cohesion stability over long periods, addressing the limitations of shorter captures found in previous datasets \cite{kolcun_revisiting_2021}. 
Accordingly, the literature motivates evaluating clustering performance with complementary internal and external metrics \cite{sawadogo_unsupervised_2022, rose_intrusion_2021, roselin_intelligent_2021}, and emphasizes the need to handle temporal drift and varying degrees of device novelty in long-duration captures \cite{rahman_unsw_2025, hao_iottfid_2023}.

\subsection{Incremental and Online Learning}

Incremental learning is essential for maintaining accuracy in dynamic IoT environments where traffic patterns drift over time \cite{kolcun_revisiting_2021,hao_iottfid_2023}. Static models often suffer performance degradation (up to 40\%) when tested on data collected outside their training period \cite{kolcun_revisiting_2021}. D-IoT is well suited to validating incremental models because it captures device behavior over an extended period under normal operating conditions and allows researchers to simulate 
temporal drift realistically \cite{pasquini_descriptor_2025}.

\subsection{Discovering Novelty and Anomalies}

Distinguishing devices with low novelty (e.g., different models from the same manufacturer) remains a primary challenge \cite{hu_network_2024,rahman_unsw_2025}. This motivates evaluating clustering under novelty conditions where new devices may be behaviorally similar to those seen during training.

\subsection{Consolidated Research Gaps}

Overall, D-IoT's long-duration, mixed IoT/non-IoT traffic supports realistic evaluation of clustering cohesion, temporal drift, and novelty conditions in a reproducible setting \cite{pasquini_descriptor_2025}. In contrast to supervised identification pipelines that depend on extensive labeled training data, our focus is on establishing strong baseline profiles without labels and then sustaining those profiles under drift and device arrival. This framing emphasizes (1) noise-aware clustering for robust initial fingerprint formation and (2) incremental updates that preserve known-device fidelity while isolating novel-device behavior with minimal retraining overhead.

\section{Methodology}

The overarching approach leverages real-world network PCAP data to extract behavioral fingerprints, employing unsupervised techniques to address the challenges of dynamic IoT environments without relying on labeled training data for initial profiling. This design aligns with established works in anomaly detection and online learning, emphasizing scalability, computational efficiency, and evaluation against ground-truth labels for validation.

\subsection{Dataset Selection}

Data selection is critical for ensuring generalization, as biased or insufficient datasets can skew unsupervised clustering results. The D-IoT dataset is selected due to its massive and comprehensive approach of collecting traffic over long periods of time from a set of devices diverse in function, make, and model.  Our experiments seeking to answer \textbf{RQ1} and \textbf{RQ2} require a deep investigation into the data and supporting files, noting devices present in the dataset, documented times for device configuration, and active and passive IoT operation windows. We analyze the D-IoT device list, noting that within the set of 24 IoT devices noted in Table \ref{tab:diot-labels}, the devices represent an excellent collection for our experiments for \textbf{RQ1} and \textbf{RQ2}, and the application of our device novelty framework.

\begin{table}[h!tb]
\caption{D-IoT Provided Device Truth Labels}
    \vspace{-0.5em}
\centering
\scriptsize
\begin{tabular}{@{}ll@{}}

\hline
\multicolumn{1}{c}{\textbf{MAC Address}} & \multicolumn{1}{c}{\textbf{Device Name}}               \\ \hline
\multicolumn{1}{|l|}{40:f6:bc:bc:89:7b} & \multicolumn{1}{l|}{Echo Dot (4th Gen)}                \\ \hline
\multicolumn{1}{|l|}{68:3a:48:0d:d4:1c} & \multicolumn{1}{l|}{Aeotec Smart Hub}                  \\ \hline
\multicolumn{1}{|l|}{70:ee:50:57:95:29} & \multicolumn{1}{l|}{Netatmo Smart Indoor Security Camera} \\ \hline
\multicolumn{1}{|l|}{54:af:97:bb:8d:8f} & \multicolumn{1}{l|}{TP-Link Tapo Pan/Tilt Wi-Fi Camera}   \\ \hline
\multicolumn{1}{|l|}{70:09:71:9d:ad:10} & \multicolumn{1}{l|}{32' Smart Monitor M80B UHD}        \\ \hline
\multicolumn{1}{|l|}{00:16:6c:d7:d5:f9} & \multicolumn{1}{l|}{SAMSUNG Pan/Tilt 1080P Wi-Fi Camera}  \\ \hline
\multicolumn{1}{|l|}{40:ac:bf:29:04:d4} & \multicolumn{1}{l|}{EZVIZ Security Camera}             \\ \hline
\multicolumn{1}{|l|}{10:5a:17:b8:a2:0b} & \multicolumn{1}{l|}{TOPERSUN Smart Plug}               \\ \hline
\multicolumn{1}{|l|}{10:5a:17:b8:9f:70} & \multicolumn{1}{l|}{TOPERSUN Smart Plug}               \\ \hline
\multicolumn{1}{|l|}{fc:67:1f:53:fa:6e} & \multicolumn{1}{l|}{Perfk Motion Sensor}               \\ \hline
\multicolumn{1}{|l|}{1c:90:ff:bf:89:46} & \multicolumn{1}{l|}{Perfk Motion Sensor}               \\ \hline
\multicolumn{1}{|l|}{cc:a7:c1:6a:b5:78} & \multicolumn{1}{l|}{NEST Protect smoke alarm}          \\ \hline
\multicolumn{1}{|l|}{70:ee:50:96:bb:dc} & \multicolumn{1}{l|}{Netatmo Weather Station}           \\ \hline
\multicolumn{1}{|l|}{00:24:e4:e3:15:6e} & \multicolumn{1}{l|}{Withings Body+ (Scales)}           \\ \hline
\multicolumn{1}{|l|}{00:24:e4:e4:55:26} & \multicolumn{1}{l|}{Withings Body+ (Scales)}           \\ \hline
\multicolumn{1}{|l|}{00:24:e4:f6:91:38} & \multicolumn{1}{l|}{Withings Connect (Blood Pressure)} \\ \hline
\multicolumn{1}{|l|}{00:24:e4:f7:ee:ac} & \multicolumn{1}{l|}{Withings Connect (Blood Pressure)} \\ \hline
\multicolumn{1}{|l|}{70:3a:2d:4a:48:e2} & \multicolumn{1}{l|}{TUYA Smartdoor Bell}               \\ \hline
\multicolumn{1}{|l|}{b0:02:47:6f:63:37} & \multicolumn{1}{l|}{Pix-Star Easy Digital Photo Frame} \\ \hline
\multicolumn{1}{|l|}{84:69:93:27:ad:35} & \multicolumn{1}{l|}{HP Envy (Printer)}                 \\ \hline
\multicolumn{1}{|l|}{18:48:be:31:4b:49} & \multicolumn{1}{l|}{Echo Show 8}                       \\ \hline
\multicolumn{1}{|l|}{74:d4:23:32:a2:d7} & \multicolumn{1}{l|}{Echo Show 8}                       \\ \hline
\multicolumn{1}{|l|}{6e:fe:2f:5a:d7:7e} & \multicolumn{1}{l|}{GALAXY Watch5 Pro}                 \\ \hline
\multicolumn{1}{|l|}{90:48:6c:08:da:8a} & \multicolumn{1}{l|}{Ring Video Doorbell}               \\ \hline
\end{tabular}
\label{tab:diot-labels}
    \vspace{-1.0em}
\end{table}

The D-IoT dataset consists of 119 PCAP files from each pcapFull and pcapIoT subset. The initial analysis generated a MAC address inventory for each PCAP file, mapped to known IoT device labels provided. This analysis informed the construction of the training and testing sets based on three categories of device novelty:
\begin{itemize}
    \item Low Novelty: Unique devices of the same make, model, and purpose as training devices.
    \item Medium Novelty: Devices of a similar purpose but different make and model compared to training devices.
    \item High Novelty: Devices of a unique make, model, and purpose, entirely unseen during training.
\end{itemize}
From this analysis, we select the Aeotec Smart Hub device as our labeled high novelty device. We identify medium novelty devices such as the cameras of multiple manufacturers and Withings devices, as well as the low novelty devices, and proceed to construct separate datasets for training and testing, drawing on files from the pcapFull subset which includes noisy background data from the thirty-six non-IoT devices. The training dataset is compiled by merging three PCAP files: \textit{2023-07-28.pcap}, \textit{2023-08-24.pcap}, and \textit{52088435\_IoT\_2023-05-19.pcap}. This composition was chosen to ensure robust representation of the known device profiles while excluding the Aeotec and one of the Withings Body+ scale devices. The \textit{2024-04-03.pcap} file was chosen for the testing dataset for its inclusion of all devices excluded from the training dataset, and was expected to provide opportunities to measure results for \textbf{RQ2}. This data-centric methodology enhances rigor, with potential extensions to larger datasets (e.g., via augmentation) for scalability. 

\subsection{Feature Selection \& Dataset Preparation}

\begin{table*}[h!tb]
    \caption{Extracted and Engineered Features}
    \vspace{-0.5em}
    \centering
    \scriptsize
    \begin{tabular}{|l|l|l|l|}
    \hline
        \textbf{Feature Name} & \textbf{Category} & \textbf{Sub-Category} & \textbf{Description} \\ \hline
        \textbf{initial\_ttl\_mode} & Static & Configuration & The statistical mode of the initial TTL observed in the IP packets of a flow. \\ \hline
        \textbf{iat\_mean} & Behavioral & Inter-Arrival Time (IAT) & Mean time between consecutive packets in the flow. \\ \hline
        \textbf{iat\_std} & Behavioral & IAT & Standard deviation of the time between consecutive packets in the flow. \\ \hline
        \textbf{iat\_median} & Behavioral & IAT & Median time between consecutive packets in the flow. \\ \hline
        \textbf{iat\_max} & Behavioral & IAT & Maximum time between consecutive packets in the flow. \\ \hline
        \textbf{total\_packets} & Behavioral & Flow Volume \& Rate & Total number of packets in the flow. \\ \hline
        \textbf{total\_bytes} & Behavioral & Flow Volume \& Rate & Total number of bytes in the flow. \\ \hline
        \textbf{flow\_duration} & Behavioral & Flow Volume \& Rate & Total duration of the flow. \\ \hline
        \textbf{packet\_rate} & Behavioral & Flow Volume \& Rate & Packet rate of the flow. \\ \hline
        \textbf{byte\_rate} & Behavioral & Flow Volume \& Rate & Byte rate of the flow. \\ \hline
        \textbf{tcp\_ratio} & Behavioral & Protocol Ratio & The ratio of TCP packets within the flow. \\ \hline
        \textbf{udp\_ratio} & Behavioral & Protocol Ratio & The ratio of UDP packets within the flow. \\ \hline
        \textbf{pkt\_size\_bin\_n} & Behavioral & Packet Size Distribution & Probability distribution of packet size for bin n where n=0..7. \\ \hline
        \textbf{top\_dst\_port\_n\_val} & Behavioral & Port Usage & The port number of the most frequently used destination ports where n=0..2. \\ \hline
        \textbf{top\_dst\_port\_n\_ratio} & Behavioral & Port Usage & The corresponding ratio of the most frequently used destination ports where n=0..2. \\ \hline
    \end{tabular}
    \label{tab:features}
    \vspace{-1.5em}
\end{table*}

Feature selection and engineering are based on prior work through processing raw network packets in PCAP files.  The processing involves identifying static and dynamic feature categories. Static features are engineered from characteristics for the device that are unlikely to change and represent device fingerprints. Dynamic features involve device behaviors created by grouping device's packets into bidirectional flows, and extracting a total of 25 numerical features for each flow. While other attributes, like DHCP hostnames, are temporarily extracted for debugging, they are excluded from the final feature set used for machine learning. This ensures the model relies purely on network behavior, not on potentially unavailable or misleading identifiers. These features are designed to capture the unique communication patterns of different IoT devices and are categorized in Table \ref{tab:features}.

With features selected, we utilize the Scapy, Pandas, and Numpy Python libraries to accomplish the data extraction then export the extracted data features into a disk cache to prevent reprocessing the large PCAP files.  This process enables efficient, iterative development of both the feature extraction process and model processing pipeline.

\subsection{Model Evaluation}

\begin{table}[h!tb]
\caption{Unsupervised Clustering Evaluation Results}
    \vspace{-0.5em}
\centering
\scriptsize
\begin{tabular}{@{}lll@{}}
\toprule
\multicolumn{1}{c}{\textbf{Model}}    & \multicolumn{1}{c}{\textbf{NMI}}       & \multicolumn{1}{c}{\textbf{Silhouette}} \\ \hline
\multicolumn{1}{|l|}{\textbf{DBSCAN}} & \multicolumn{1}{l|}{\textbf{0.780009}} & \multicolumn{1}{l|}{0.925894}           \\ \hline
\multicolumn{1}{|l|}{HDBSCAN}       & \multicolumn{1}{l|}{0.651134} & \multicolumn{1}{l|}{0.865030}          \\ \hline
\multicolumn{1}{|l|}{K-Means}       & \multicolumn{1}{l|}{0.021334} & \multicolumn{1}{l|}{\textbf{0.994177}} \\ \hline
\multicolumn{1}{|l|}{BIRCH}         & \multicolumn{1}{l|}{0.021303} & \multicolumn{1}{l|}{0.994043}          \\ \hline
\end{tabular}
\label{tab:ucm-results}
    \vspace{-1.5em}
\end{table}

\subsubsection{Cluster Cohesion Metrics}
To evaluate the efficacy of unsupervised learning models, researchers rely on specific metrics that assess the structural quality of clusters and their alignment with ground truth data. The Silhouette Coefficient is an internal validation metric that quantifies the quality of clustering by measuring both cluster cohesion and separation \cite{sawadogo_unsupervised_2022}. Conversely, Normalized Mutual Information (NMI) is a standard external evaluation metric used to validate clustering accuracy when ground-truth labels are available \cite{rose_intrusion_2021}. NMI values range from 0 to 1, where a higher score indicates a stronger correlation between the predicted clusters and the actual device types, implying more accurate clustering results \cite{roselin_intelligent_2021}. In descriptive terms, we consider NMI as a measure of cluster purity (alignment with truth labels) and the Silhouette Coefficient as a measure of cluster density and separation. However, the D-IoT dataset's inclusion of 19 distinct device types and 36 non-IoT devices creates a more challenging and realistic feature space for evaluating separation metrics like cluster cohesion (Silhouette Coefficient) and NMI, particularly when separating functionally similar devices \cite{pasquini_descriptor_2025}.

\subsubsection{Benchmark Model Selection}
Prior to performing experiments, given that we are not reproducing prior work related to unsupervised clustering models, we evaluate the classic clustering models against the dataset with an iterative, grid-search process to tune parameters seeking to select the best unsupervised clustering model to address \textbf{RQ1}. The results of this experiment shown in Table \ref{tab:ucm-results} require deeper analysis.

\begin{figure}[h!tb]
\centering
\begin{subfigure}{0.4\textwidth}
    \includegraphics[width=0.8\columnwidth]{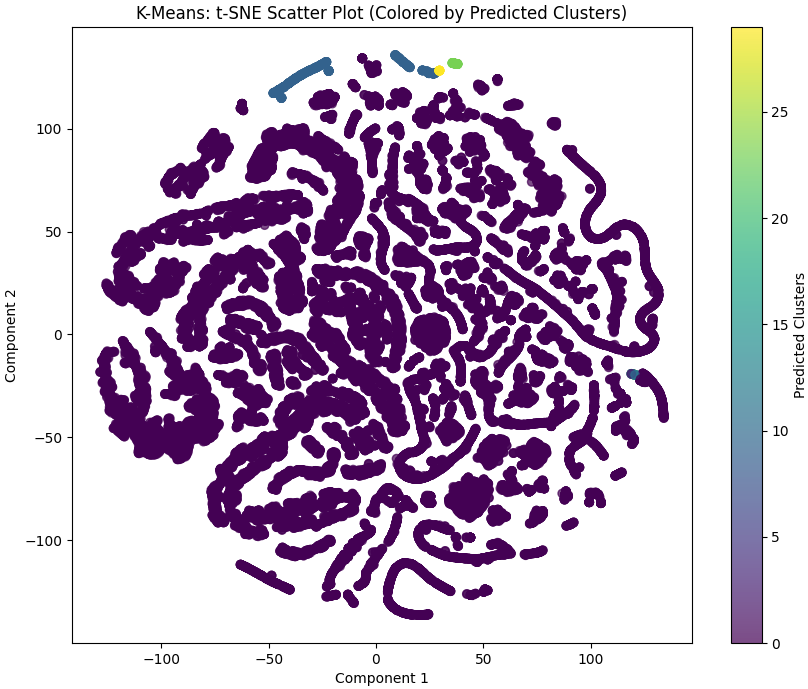}
    \vspace{-0.5em}
    \caption{K-Means: near-zero NMI \& high Silhouette}
    \label{fig:kmeanseval}
\end{subfigure}
\begin{subfigure}{0.4\textwidth}
    \includegraphics[width=0.8\columnwidth]{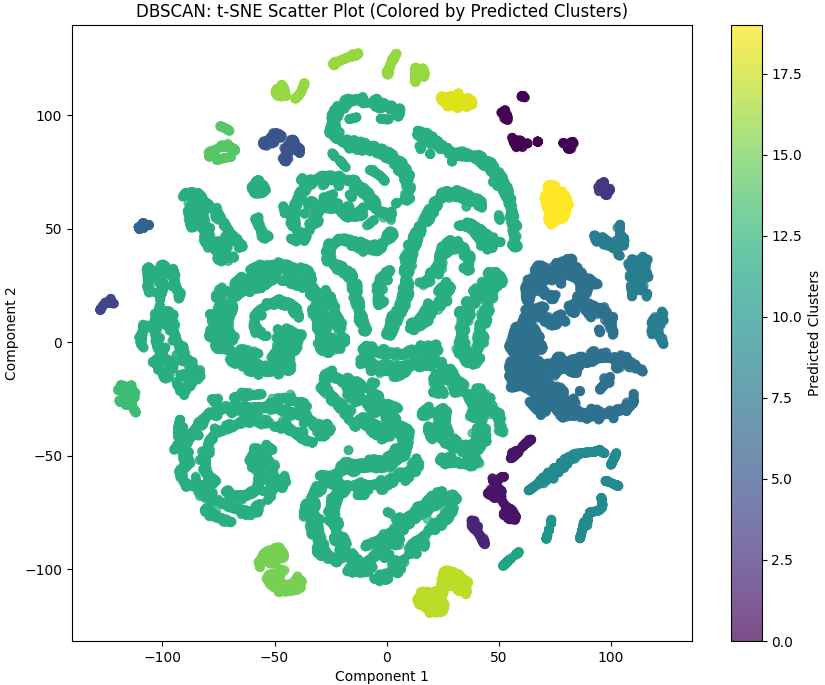}
    \vspace{-0.5em}
    \caption{DBSCAN: high NMI \& lower Silhouette}
    \label{fig:dbscaneval}
\end{subfigure}    
\caption{t-SNE 2D Cluster Visualizations}
\label{fig:eval}
    \vspace{-1.0em}
\end{figure}

\textbf{NMI} and \textbf{Silhouette Coefficient} scores are measured, resulting in inconclusive results to determine the best model by the score alone. Further analysis indicates that NMI carries greater importance for our goals in \textbf{RQ2} and should be the principal metric while also trying to maximize the \textbf{Silhouette} score. NMI represents an external metric measured with the true labels for our data, making it an ideal measure to understand the \textit{purity} of data points within a cluster.  Conversely, \textbf{Silhouette} represents an internal measure that doesn't rely on truth labels and this signifies that clusters are \textit{distinct} and well defined, but will not reflect cluster \textit{purity}.  We aim for cluster \textit{purity} when attempting to form clusters around device profiles. This analysis leads to the selection of DBSCAN because it has earned the top NMI score while maintaining a respectable Silhouette score as shown in Table \ref{tab:ucm-results}.  The inability for \textbf{Silhouette} scoring to stand alone is seen visually represented with t-SNE 2 dimensional cluster representations of K-Means in Fig. \ref{fig:kmeanseval} with near-zero NMI but very high Silhouette scores revealing almost all data falls into a single cluster.  Conversely, the same visualization for DBSCAN in Fig. \ref{fig:dbscaneval} with high NMI and medium Silhouette scores revealing many strong, well defined clusters. This benchmarking phase, while exploratory, establishes DBSCAN's baseline efficacy, revealing its superiority over centroid-based methods like K-Means, which falter due to assumptions of spherical clusters that do not hold in high-dimensional network feature spaces.

\subsection{Experiment Methodology}
This study employs a multi-stage machine learning pipeline designed to identify and profile IoT devices using unsupervised clustering and incremental learning. The methodology addresses the challenge of dynamic IoT environments where ground truth labels are unavailable and new devices are frequently introduced by taking quantitative measures of purity and cohesion from unsupervised learning methods as well as computational performance to address the research questions. 

Prior studies have indicated excellent NMI results from deep learning methods combined with BIRCH that exceed our evaluation results, however Silhouette is not taken into account \cite{roselin_intelligent_2021}. Similarly, previous studies have approached the problem with classical learning methods achieving respectable Silhouette scores, but not measuring NMI \cite{sawadogo_unsupervised_2022,kumar_s_comparative_2023}.  Other researchers have approached the problem using more traditional scoring such as Accuracy, Precision, Recall, and F1, but did not measure NMI or Silhouette scores \cite{deng_iot_2025}. Our methodology seeks to build a model which can optimize both measures for answering \textbf{RQ1} and \textbf{RQ2}, and provide indicators of both well defined boundaries and strong cluster purity.

To aid the iterative model development process for \textbf{RQ2}, two metrics serve as analogs to Precision and Recall, labeled as \textit{Purity} and \textit{Share}, respectively. Purity measures how ``clean'' the clusters associated with the new device are (i.e., whether they contain only that device's traffic or are ``polluted'' by known-device traffic) and is calculated as a weighted average. Share measures recall or ``capture rate.'' 
It answers the question: ``Of all the traffic generated by this new device, what percentage end up in high-quality, distinct clusters?''

$$ Purity = \frac{\sum_{c \in C_{valid}} \left( \frac{n_{novel, c}}{n_{novel, c} + n_{known, c}} \times n_{novel, c} \right)}{\sum_{c \in C_{valid}} n_{novel, c}} $$

Where:
\begin{itemize}
    \item $C_{valid}$ is the set of subclusters where the local purity meets the defined threshold.
    \item $n_{novel, c}$ is the count of novel device samples in subcluster $c$.
    \item $n_{known, c}$ is the count of known device samples in subcluster $c$.
\end{itemize}

$$ Share = \frac{\sum_{c \in C_{valid}} n_{novel, c}}{N_{total}} $$

Where:
\begin{itemize}
    \item $C_{valid}$ is the set of subclusters where the local purity meets the defined threshold.
    \item $n_{novel, c}$ is the count of novel device samples in subcluster $c$.
    \item $N_{total}$ is the total number of samples belonging to the novel device in the evaluation set.
\end{itemize}

\begin{table*}[h!tb]
\caption{Summary of RQ1 (static baseline) and RQ2 (incremental adaptation) evaluation metrics.}
\label{tab:rq-summary}
\vspace{-0.5em}
\centering
\scriptsize
\begin{tabularx}{\textwidth}{@{}l *{8}{>{\centering\arraybackslash}X} @{}}
\toprule
\textbf{Setting / Model} &
\textbf{Clusters} &
\textbf{Noise \%} &
\textbf{NMI} &
\textbf{Silhouette} &
\textbf{Known Acc. (Post)} &
\textbf{Novel Purity} &
\textbf{Novel Share} & 
\textbf{Update Time (s)}\\
\midrule
RQ1 (Static) DBSCAN & 20 & 41.21 & 0.7799 & 0.9237 & -- & -- & -- & -- \\
\midrule
RQ2 (Incremental) MiniBatchKMeans & -- & -- & 0.4434\textsuperscript{1} & 0.0954\textsuperscript{1} & 0.6922 & 0.0000 & 0.0000 & 0.0011 \\
RQ2 (Incremental) BIRCH & -- & -- & 0.4292\textsuperscript{1} & 0.6797\textsuperscript{1} & 0.7121 & 0.8664 & 0.7240 & 0.1337\\
\bottomrule
\multicolumn{8}{@{}l@{}}{\footnotesize \textsuperscript{1}For RQ2, NMI and Silhouette are reported globally over known+novel traffic after adaptation.}\\

\end{tabularx}
\vspace{-0.75em}
\end{table*}
We conduct the \textbf{RQ1} experiment by reproducing the results of the exploratory model evaluation.  A new DBSCAN model is created using the parameters discovered by grid search in the evaluation. \textbf{RQ1} computes Silhouette and NMI scores after filtering noise points. Additional outputs include a 2D PCA visualization of clusters and a membership table, providing qualitative corroboration of quantitative metrics. For instance, the high NMI (0.7800) observed in benchmarking is reaffirmed, with noise handling aligning precisely, as DBSCAN's density-based mechanism identifies outliers without compromising cluster integrity. Retraining DBSCAN under these conditions yields identical or negligible variations in scores due to stochastic elements. This consistency validates DBSCAN's robustness for \textbf{RQ1}, where unsupervised profiling must reliably segregate device types (e.g., cameras, smart plugs) based on traffic patterns alone. The emphasis on noise exclusion ensures metrics reflect meaningful clusters, avoiding inflation from poorly defined groups. By reproducing high NMI and Silhouette scores, the pipeline affirms the model's ability to profile known IoT devices without labels, aligning with principles of scalability in network analysis. However, it also highlights limitations for subsequent research questions: DBSCAN's static nature necessitates alternatives for incremental adaptation in \textbf{RQ2}, as its lack of online learning support impedes novelty detection without full retraining. Future enhancements might involve integrating feedback loops or hybrid models to extend these validated results.

To answer \textbf{RQ2}, an incremental learning model is added to the pipeline.  Given that DBSCAN is not suitable for this process, prior studies \cite{roselin_intelligent_2021,sawadogo_unsupervised_2022} motivate investigating MiniBatchKMeans (a K-Means variant) and BIRCH for use in the incremental pipeline. MiniBatchKMeans was selected for its ability to process mini-batches of data incrementally, aligning with \textbf{RQ2}'s focus on adaptation costs (e.g., computational time for model updates) and novelty detection. However, empirical results revealed significant shortcomings. MiniBatchKMeans, like standard K-Means, assumes spherical clusters and relies on Euclidean distance, which poorly accommodates the non-spherical, high-dimensional feature spaces derived from IoT network traffic. Evaluations demonstrated low NMI scores (e.g., ~0.0213 for K-Means), indicating misalignment with ground-truth labels despite high Silhouette coefficients (reflecting tight but misaligned clusters). In \textbf{RQ2}, this translated to suboptimal adaptation: the model struggled with catastrophic forgetting and inadequate novelty handling, as novel devices often forced reshuffling of centroids without clear separation. Recognizing these deficiencies, the implementation pivoted to BIRCH, a hierarchical clustering algorithm optimized for incremental learning and large-scale data streams. BIRCH constructs a Clustering Feature (CF) tree that summarizes data in subclusters, enabling efficient updates without recomputing the entire model, a key advantage for \textbf{RQ2}'s adaptation requirements. This shift yielded notable improvements over MiniBatchKMeans. BIRCH's hierarchical approach better handles fragmentation and noise in IoT features.

\section{Results and Analysis}

To make the quantitative comparison across static (RQ1) and incremental (RQ2) settings explicit, Table~\ref{tab:rq-summary} summarizes the resulting evaluation metrics. The results include NMI (external alignment / cluster purity) and Silhouette (internal cohesion) for cluster quality, along with noise rate (RQ1) and adaptation trade-off metrics (RQ2). Raw console logs and auxiliary diagnostic artifacts (e.g., intermediate cluster membership tables) are omitted for brevity.

\subsection{Analysis}

\begin{figure}[h!tb]
\centering
\begin{subfigure}{0.4\textwidth}
    \includegraphics[width=0.8\columnwidth]{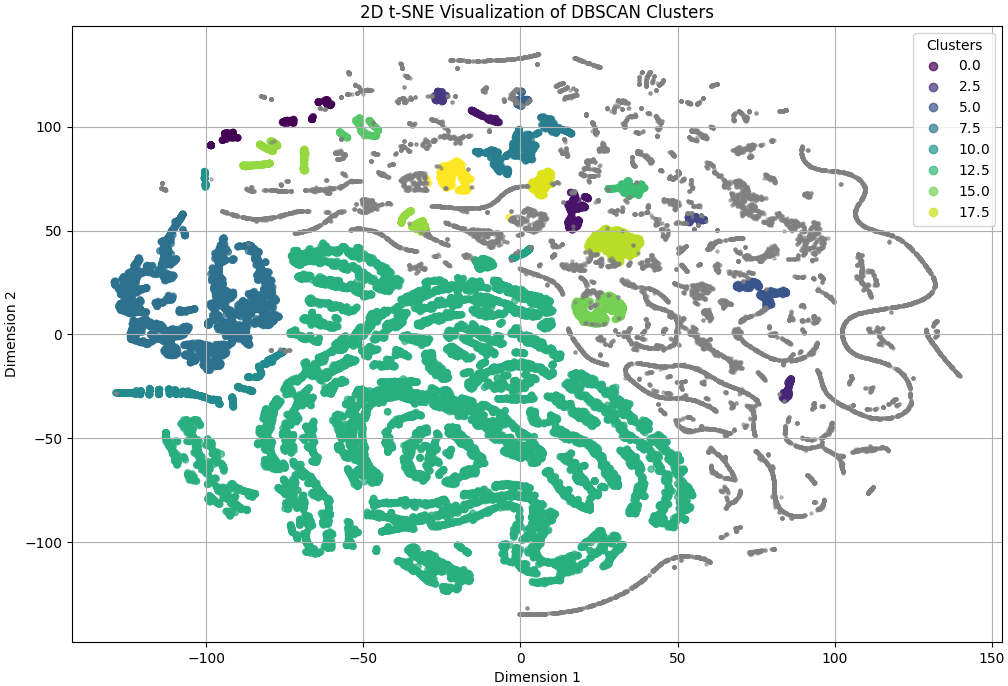}
    \vspace{-0.5em}
    \caption{RQ1: DBSCAN}
    \label{fig:rq1vis}
\end{subfigure}
\begin{subfigure}{0.4\textwidth}
    \includegraphics[width=0.8\columnwidth]{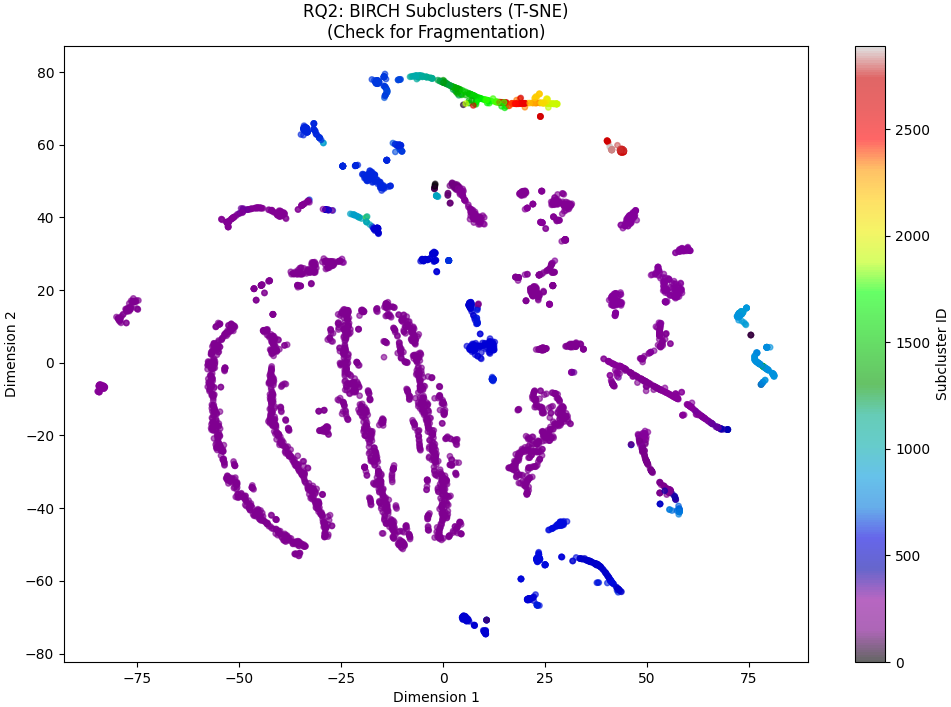}
    \vspace{-0.5em}
    \caption{RQ2: BIRCH}
    \label{fig:rq2vis}
\end{subfigure}    
\caption{t-SNE 2D Cluster Visualizations}
\label{fig:rqvisuals}
    \vspace{-1.0em}
\end{figure}

A comparison of cluster visualizations for \textbf{RQ1} and \textbf{RQ2} are provided in Fig. \ref{fig:rqvisuals}. BIRCH does not surpass DBSCAN's performance in \textbf{RQ1}, highlighting fundamental differences between static unsupervised clustering and incremental adaptation. DBSCAN achieves superior NMI (0.7800) by leveraging density-based connectivity to form coherent clusters that align closely with ground-truth labels (e.g., distinguishing cameras from smart plugs). Its noise-handling (e.g., excluding ~41\% outliers) and parameter tuning (via k-distance plots) optimize for known-device profiling without adaptation overhead. In contrast, \textbf{RQ2}'s incremental models prioritize adaptability over initial accuracy. BIRCH's subcluster-based approach, while effective for novelty detection often results in lower global NMI (as seen in \textbf{RQ2} results) due to fragmentation or label mapping challenges. For example, trade-offs like reduced known-device accuracy post-adaptation and variable cluster purity/share for novel devices are observed, reflecting the complexity of maintaining fidelity to known profiles while integrating new ones. This disparity is not a failure but a design trade-off: DBSCAN excels in static, noise-tolerant profiling (\textbf{RQ1}), whereas BIRCH enables dynamic evolution (\textbf{RQ2}), at the cost of some discriminative power. No single model dominates both domains, underscoring the need for hybrid approaches in future iterations.

\section{Conclusion}
This study evaluated an adaptive, packet-efficient approach to IoT device profiling in dynamic network environments, where new devices must be profiled without extensive labeling or full retraining. Using real-world PCAP traffic from a diverse set of IoT devices, results show that a two-stage strategy, i.e., strong baseline unsupervised profiling followed by incremental adaptation, can support practical device fingerprinting in evolving deployments. 

For baseline profiling (RQ1), density-based clustering (DBSCAN) produced the most label-aligned device groupings and remained robust to background noise, making it a strong choice for establishing initial device fingerprints from behavioral flow features.  In contrast, incremental adaptation (RQ2) prioritizes continual evolution of profiles as novel devices appear. Our results indicate that BIRCH is better suited than centroid-based incremental clustering for this setting, enabling efficient updates and novelty clustering, but with an expected reduction in global cluster coherence compared with the best static profiler, reflecting the core trade-off between discriminative power and online flexibility. Collectively, these findings support a pragmatic pathway from unsupervised device fingerprint formation to incremental refinement for long-running IoT networks, without requiring deep learning methods that may be impractical for constrained deployments. 

Key limitations remain. First, DBSCAN is not directly incremental, while BIRCH can fragment into subclusters under highly variable traffic, complicating label mapping and reducing global coherence. Second, the approach assumes access to packet header/flow-level metadata (e.g., sizes, timing, ports); deployments that expose only coarse aggregates, or that heavily tunnel/obscure metadata, may reduce separability even if payloads are not used. Third, although incremental update time for adaptation is reported, a fully retrained-from-scratch baseline for periodic retraining is not included. Finally, while the dataset spans long durations, the evaluation uses fixed training/testing windows and does not quantify behavior drift across many timepoints (e.g., rolling-window degradation curves). Future work should therefore focus on hybrid designs that preserve DBSCAN-quality baseline structure while enabling BIRCH-style adaptation, and include periodic full-retraining baselines, rolling drift evaluation, and richer temporal features to improve generalization under long-term behavioral change.

\section{Acknowledgments}
ChatGPT assisted with grammar and organization. The content, analysis, and ideas are the authors' original work.
\printbibliography
\end{document}